\documentclass{appolb}
\usepackage{epsfig}
\begin{document} 
\eqsec  
\title{Thermal bremsstrahlung probing the nuclear liquid-gas phase transition 
\thanks{Presented at TAPS Workshop VI, Krzy\.ze, Poland, September 9-13, 2001}}
\author{R. Ortega  
\address{Grup de F\'{\i}sica de les Radiacions, Universitat Aut\` onoma de
Barcelona 08193 and SUBATECH, 4, rue Alfred Kastler BP20722 44307
Nantes Cedex 3, France}
\and
 F. Fern\'andez 
\address{Grup de F\'{\i}sica de les Radiacions, Universitat Aut\` onoma de
Barcelona 08193}, \\
D. d'Enterria and G. Mart\'{\i}nez 
\address{SUBATECH, 4, rue Alfred Kastler BP20722 44307
Nantes Cedex 3, France}
\\for the TAPS collaboration}
\maketitle
\begin{abstract}
We present the results of the analysis of the hard photon production in the  $^{129}${Xe}+$^{\rm nat}${Sn} 
at 50{\it A} MeV system studied in the GANIL E300 experiment. The  energy and angular hard
photon distributions confirm the existence of a thermal component which follows the recently  measured
thermal bremsstrahlung systematics. Exploiting the performances of our complete detection system, 
consisting of TAPS and 3 charged particle multidetectors, we have also  measured the hard photon multiplicity 
as a function of the charged particle multiplicity.
\end{abstract}
  
\section{Introduction}
 One of the main issues  in heavy-ion reactions at intermediate bombarding energies, from a few tens
 up to a hundred {\it A}MeV,  is to determine the origin of  nuclear multifragmentation. This process is
 characterized by
 an abundant production of intermediate-mass-fragments (3 $\leq$ Z $\leq$ 20) and it is the dominant
 decay channel observed in central nuclear collisions. The interest of multifragmentation relies on 
 its possible connection with the expected, though elusive, liquid-gas phase transition of nuclear
 matter \cite{more,gupta}. 
 As a matter of fact, the origin and dynamics 
 of multifragmentation are  not yet clearly pinned down. On one extreme, sequential statistical approaches
 sustain that multifragmentation is the result of  a slow and sequential fragment production  process starting 
 from a thermally equilibrated source \cite{fri}. On the other extreme, purely dynamical models
 assert that the fragmenting system has not  reached global equilibrium when it  disassembles in a prompt 
 process \cite{bal,aic}. In between those extremes, fast statistical models \cite{bot,gross} rely on the
 attainment of (local) thermal equilibrium followed by a fast expansion of the hot nuclear systems which then
 break up according to the available phase space. All these seemingly different models reproduce well several
 patterns observed in  the fragment production data. Therefore, to disentangle between them and shed some
 light on the connection between multifragmentation and the nuclear liquid-gas phase transition, one would
 need a new experimental observable contemporary to multifragmentation. We argue that thermal hard photons
 can be such a probe \cite{dav00}.
    
   Hard photons are bremsstrahlung photons of energies above 30 MeV produced in proton-neutron collisions.
 Most part of the hard photon emission, the direct one,
 is produced in the initial preequilibrium stage of the reaction during the first compression of the system, 
 but there is a softer
 second emission corresponding to the so-called thermal hard
 photons, which are emitted at intermediate stages of the reaction \cite{gines,yves}. Recently, the analysis
 of the data taken in Ar-induced reactions at 60{\it A} MeV by the TAPS collaboration at the Dutch-French
 AGOR cyclotron at KVI, has demonstrated that this second hard photon component really emerges from a
 thermalized source \cite{dav00}
 during the time scale when the multifragmentation process is
 supposed to take place (50 fm/c\,-\,200 fm/c). Moreover, the measurement of a thermal hard photon emission in 
 multifragmentation reactions in the  $^{36}$Ar+$^{197}$Au system has demonstrated that, at least for this
 reaction, the source breaks up after thermalization (t\,$\geq$ 80 fm/c) \cite{davt,dav00}. The observation of
 a thermally equilibrated radiating nuclear source opens up an interesting possibility in the study of the 
 equation of state of nuclear matter in the region of the predicted liquid-gas phase transition \cite{davnew}. 

 The aim of the experiment E300 performed at GANIL in 1998  is to 
 investigate the hard-photon production in multifragmentation reactions of the   $^{129}${Xe}+$^{\rm nat}${Sn} 
 at 50{\it A} MeV system \cite{ginespol}. The characteristics of this projectile-target combination (a large
 quasi-symmetric system with large energy deposition in the center-of-mass) makes of it a good candidate to
 study the liquid-gas transition. The  existence/absence of a thermal 
 hard-photon emission in multifragmentation reactions in such a system is expected to deliver, by means of
 a comparison with the previous KVI experimental results, a more definite conclusion of the origin of 
 nuclear multifragmentation.
 \section{Experimental set-up}
 To carry out this investigation inclusive and  exclusive charged particle - photon measurements
 are necessary.
\begin{figure}
 \begin{center}
 \mbox{\epsfxsize=10cm  \epsfbox{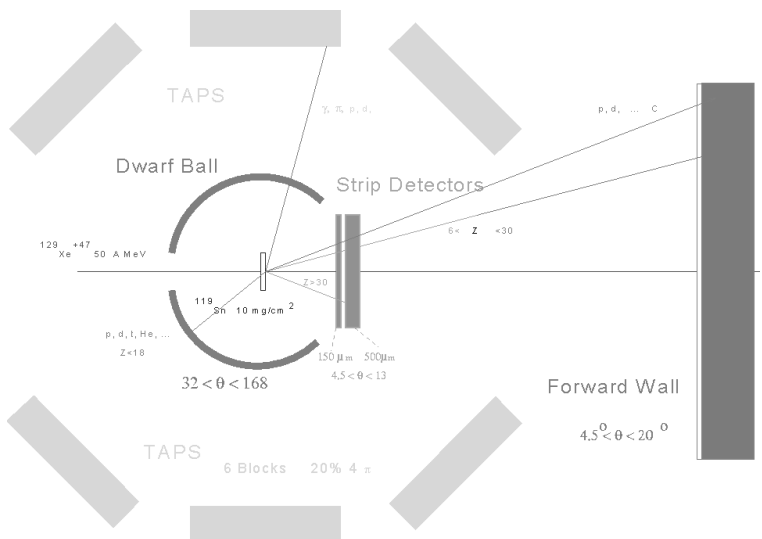}}
 \end{center}
 \caption{Schematic view of the experimental setup used to study the reaction 
 $^{129}${Xe}+$^{nat}${Sn} at 50{\it A} MeV. The detector system  consisted of 
 the ``Silicon Strip Detector" (SSD), the ``Dwarf Ball" (DB), the TAPS photon
 spectrometer and the 
 ``Forward Wall"(FW.)}
 \label{fig: det.eps}
\end{figure}
 For that purpose, the experimental setup of the E300 experiment (Fig. 1) consisted of the TAPS photon 
 spectrometer coupled for the first time with
 three different charged-particle multidetectors:  the ``Silicon Strip Detector" (SSD), the Washington
 University ``Dwarf Ball" (DB) and the KVI ``Forward Wall"(FW). TAPS \cite{nov},  arranged in 6
 blocks of 64 BaF$_2$ each, allowed  to measure photons  of 5 MeV$\,<\mbox{
 E$_\gamma$}<\,$200 MeV in  20\,\% of the full solid angle. The SSD \cite{jos}, consisting of 64 circular and 128 radial 
 silicon detectors, covered a forward angular range of $2.0^\circ \leq
 \theta \leq 10.3^\circ$. The SSD is sensitive to the intermediate-mass-fragments (IMF) and
 projectile-like-fragments (PLF) emitted at forward angles.  The DB \cite{str} is a  multidetector 
 consisting of 64 (BC400-CsITl) phoswich detectors forming a sphere surrounding the target and covering
 and angular range of $32^\circ \leq \theta \leq 168^\circ$. The DB permitted the identification of  
 light-charged-particles (LCP) and IMF. The FW \cite{luck} is 
 also a phoswich multidetector, with 92 (NE115-NE102A) plastic modules, which allows the identification of
 LCP and IMF. The FW was placed downstream from the target covering the forward hemisphere 
 ($2.5^\circ \leq \theta \leq 25.^\circ$). The energy deposited by charged particles in the circular and radial strips of the SSD and in
 the individual  crystal and plastic of each DB and FW  phoswich were recorded. The charge of detected
 fragments and particles is identified by $\Delta$E versus E analysis. The full details of  the experimental
 setup and off-line analysis have been reported elsewhere \cite{raqrez,raq}.

\section{Experimental  results}
\subsection{Inclusive hard photon results}
   After applying an energy correction factor to correct energy calibration and energy losses in the BaF$_2$ modules
 and subtracting the cosmic background we obtain the energy and angular spectra  of the inclusive, i.e. without 
 requiring any specific exit-channel condition, photon measurements \cite{raq}. 

 In agreement with previous TAPS experiments we have found that the hard-photon energy
 spectrum  measured in the NN center-of-mass (Fig. 2.a) is well reproduced  with a double source fit, consisted
 of a sum of two exponential distributions: 
\begin{equation}
 \label{eq:2exponentials}
 \frac{dN}{d E_\gamma}\,=\,K_d\:e^{-E_\gamma/E_0^d}+K_t\:e^{-E_\gamma/E_0^t}
\end{equation}

\begin{figure}[htbp]
 \begin{center}
 \mbox{\epsfxsize=6.5cm  \epsfbox{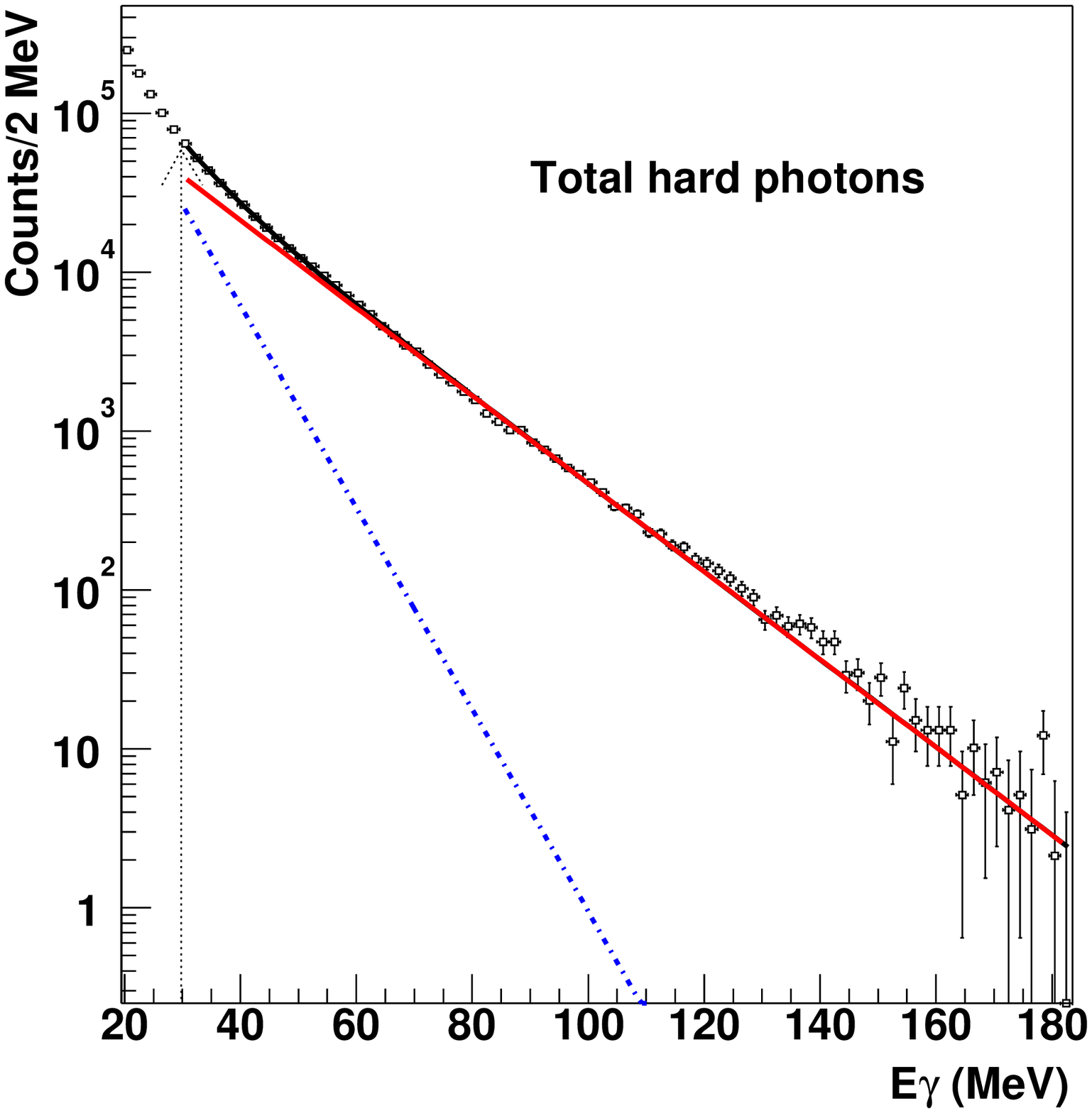} \epsfxsize=6.8cm  \epsfbox{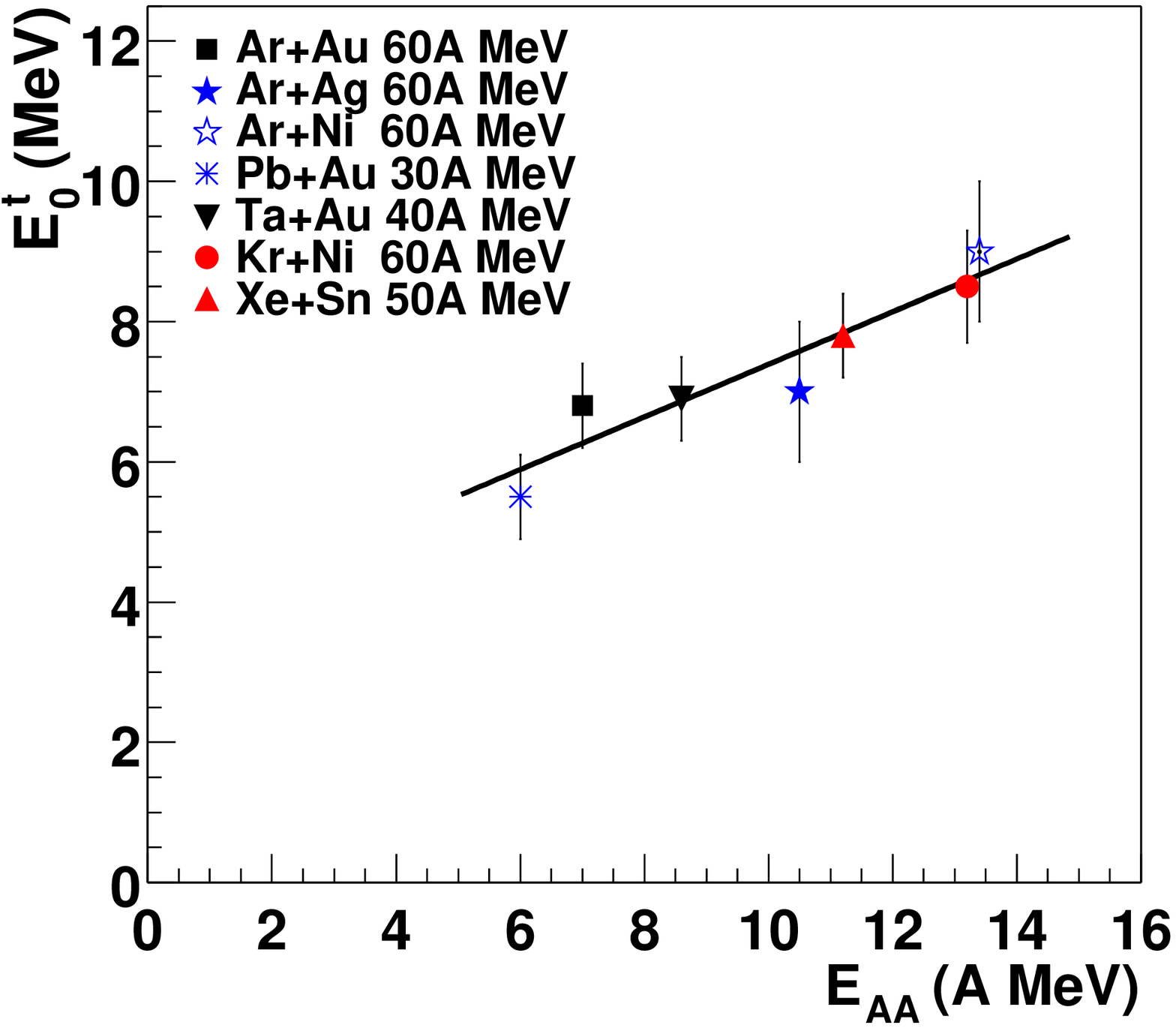}}
\end{center}
\caption{a) The experimental inclusive hard photon spectrum measured for the reaction $^{129}${Xe} +
 $^{112}${Sn} in the NN center-of-mass frame fitted to Eq. (3.1).  The thermal (dashed
 line) and direct (solid line) exponential contributions are shown. b) Thermal hard-photon slopes $E_0^t$, 
 measured at $\theta_\gamma^{lab}\,=\,90^\circ$, plotted
 as a function of the Coulomb-corrected
 nucleus-nucleus center-of-mass energy $E_{Cc}^{AA}$. The  
 measurements correspond to different TAPS experiments, the up triangle denotes the value of the reaction
 described in this contribution. The solid line is a linear fit to the
 data. Figure extracted from \cite{dav00}.}
\label{fig: en.eps}
\end{figure}
 where the $K_{d,t}$ factors are defined by the direct and thermal hard-photon
 intensities, respectively.
 The  harder hard-photon distribution with slope  $E_0^d$\,=\,15.6 $\pm$ 1.0 MeV corresponds to the dominant
 direct  component, whereas the softer one ($E_0^t$\,=\,7.0 $\pm$ 0.6 MeV)
 accounts for the thermal contribution, which amounts to 22\% of the total intensity. Above 
 $E_\gamma>60$ MeV only the direct hard-photon emission is significant.
 The measured thermal slope, $E_0^t$, follows the recently observed linear dependence on the available energy in
 the nucleus-nucleus center-of-mass, $E_{Cc}^{AA}$ (Fig. 2.b) \cite{dav00}. This observation is consistent
 with a thermal bremsstrahlung emission from secondary $pn$ collisions.   The obtained  experimental thermal
 hard photon multiplicity
  $M_\gamma^{exp}$\,=\,$\sigma_\gamma^{exp}$\,/\,$\sigma_R^{exp}$\,=\,(2.6 $\pm$ 0.4)$10^{-4}$.
\begin{figure}
 \begin{center}
 \mbox{\epsfxsize=6.7cm  \epsfbox{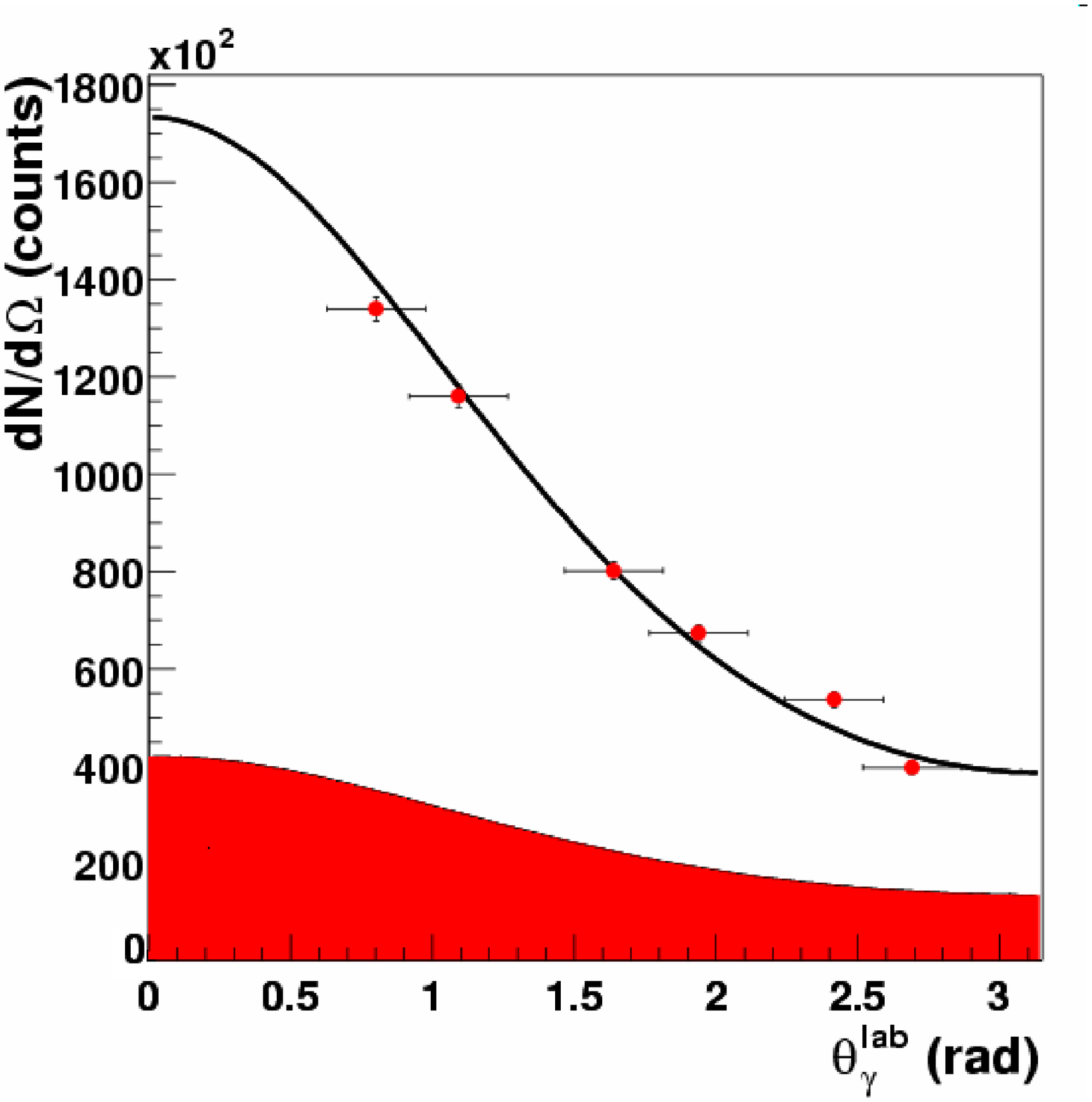}}
 \end{center}
\caption{Hard photon ($E_\gamma>\,30$ MeV) angular distribution measured in the laboratory frame. The dark
 region is an estimation of the thermal hard-photon contribution.}
\label{fig: ang:tot.eps}
\end{figure}

 We have analyzed the hard-photon angular distribution in  two
 different energy ranges: the region of photons with 30 MeV$\,<\mbox{ E$_\gamma$}<\,$40 MeV and that of 
$E_\gamma>\,60$ MeV. The first energy region   consists of  a mixed
 distribution of direct and thermal photons and the latter exhibit their largest intensity,
 between 38\% and 29\% of the total hard photon yield. Whereas, as aforementioned, for  $E_\gamma>\,60$ MeV   direct 
 hard-photons clearly dominate.  The measured (Doppler-shifted) laboratory angular
 distributions integrated over the two energy ranges (30 MeV$\,<\mbox{ E$_\gamma$}<\,$40 MeV, and
 $E_\gamma>\,60$ MeV) have been respectively fitted according to the following expressions:
 \begin{equation}
 \label{eq:lab angular distribution 30-40}
 \left(\frac{dN}{d\Omega}\right)_{lab}\,=\,\frac{K}{Z^2}\left[1-\alpha+\alpha
 \,\frac{\sin^2\theta_{\gamma}^{lab}}{Z^2}\right]\,E_0\,
\left[e^{-30\cdot Z/E_0}-e^{40\cdot Z/E_0}\right]
\end{equation}
 \begin{equation}
 \label{eq:lab angular distribution 60}
\left(\frac{dN}{d\Omega}\right)_{lab}\,=\,\frac{K}{Z^2}\left[1-\alpha+\alpha
\,\frac{\sin^2\theta_{\gamma}^{lab}}{Z^2}\right]\,E_0\,e^{-60\cdot Z/E_0}
\end{equation}

 where $Z\,=\,\gamma_{S}(1-\beta_{S}\cos\theta_{lab})$, K is a
 normalization factor, $\beta_{S}$ is
 the source velocity, $\alpha$ is the anisotropy factor and $E_{0}$ is the local slope.\\
 The  velocity of the hard-photon source  obtained in both fits is nearly the 
 same, $\beta_{S}\,=\,0.16\,\pm$\,0.01, and it agrees  with the emission from a source moving with
 $\beta_{S}\,=\,\beta_{NN}$\ but also, due to the symmetry of the
 $^{129}${Xe}+$^{\rm nat}${Sn} system, with $\beta_{S}\,=\,\beta_{AA}$. Due to this 
 symmetry,  we  have hence one free parameter less in the analysis of the angular distribution and
  therefore we are more sensitive to the fitting value  of the anisotropy factor. The anisotropy factor  
  is due to the existence of a preferential direction of the dipole component  of the bremsstrahlung
   radiation in the elementary pn$\rightarrow$pn$\gamma$ process. This anisotropy should only 
 appear in the radiation emerging from first-chance NN scattering, in which the momentum of the 
 projectile nucleon still conserves its original (beam) direction. Indeed, we observe a non zero
 anisotropy factor, $\alpha\,=\,0.2\,\pm$\,0.1, in the angular distribution measured for photons of 
 $E_\gamma>\,60$ MeV, which is  dominated by the direct hard-photon component. In contrast, no anisotropy is
  found in the angular distribution of the mixed  thermal and direct energy range
 (30 MeV$\,<\mbox{ E$_\gamma$}<\,$40 MeV): $\alpha\,=\,0.0\,\pm$\,0.1. 
 Thus, these slightly different (within  their associated fitting errors) $\alpha$ values can be 
 interpreted as an indication of the distinct origins of the two hard-photon contributions. 
 Finally, the  total ($E_\gamma>\,30$ MeV) hard photon angular emission (Fig. 3) can be well 
 reproduced with the expression:
\begin{equation}
 \label{eq:lab angular distribution 30-200}
\left(\frac{dN}{d\Omega}\right)_{lab}\,=\,\frac{K^{d}}{Z^2}\left[1-\alpha+\alpha\,
\frac{\sin^2\theta_{\gamma}^{lab}}{Z^2}\right]\,E_0^{d}\,e^{-30\,Z/E_0^{d}}\,
 + \frac{K^{t}}{Z^2}\,E_0^{t}\,e^{-30\,Z/E_0^{t}}
\end{equation}

The values of the direct/thermal and the ratio of intensities are those measured in the
energy spectrum.
\subsection{Exclusive hard photon measurements}
 In our previous analysis we have  focused on the global properties of the hard photon emission. However, if 
 we want to shed light on the mechanism leading to multifragmentation and to extract the  thermodynamical 
 properties of a possibly thermalized photon source,  we have to select  the more interesting (though less probable) 
 central and semi-central reactions, and analyze the
 hard photon production emitted in such collisions. LCP and IMF multiplicities will allow us to define 
 the impact-parameter selection criteria. Next, we present our first exclusive photon measurements: 
 the correlation of photon multiplicities for  3 energy ranges with the charged particle 
 multiplicities measured in the DB and FW detectors. We will consider three energy regions: 10 MeV$\,<\mbox{
 E$_\gamma$}<\,$22 MeV and hard-photons of  30 MeV$\,<\mbox{ E$_\gamma$}<\,$40 MeV and  
 $E_\gamma>\,60$ MeV. The two regions of hard-photons have been chosen
 due to the reasons explained in the analysis of the angular distribution. Photons of energies between 10-20 
 MeV come mainly from the decay of  Giant-Dipole Resonances (GDR) produced in the moderated excited fragments and,
 unlike hard photons,  are produced by a collective mechanism.  As Fig. 4.a shows, one can separate the two hard photon 
 components from GDR photons by  the  dependence of their  multiplicities  on the DB charged particle 
 multiplicity. Whereas both hard photon components increase with the charged particle
 multiplicity, i.e with the centrality, GDR photons exhibit a maximum  at peripheral reactions. Fig. 4.b shows
 the same histogram of M$_{\gamma}$ versus M$^{DB}_{CP}$ with the additional  condition on 
 the charged particle multiplicity measured in the FW:   
 M$^{FW}_{CP}$=2-4, where   the photon multiplicity exhibits a maximum. In this case, we 
 observed the maximum  enhancement of hard-photons  with  charged particle multiplicity: a factor 10 between 
 M$^{DB}_{CP}$=1
 and M$^{DB}_{CP}$=11. 
\begin{figure}
 \begin{center}
 \mbox{\epsfxsize=7.0cm  \epsfbox{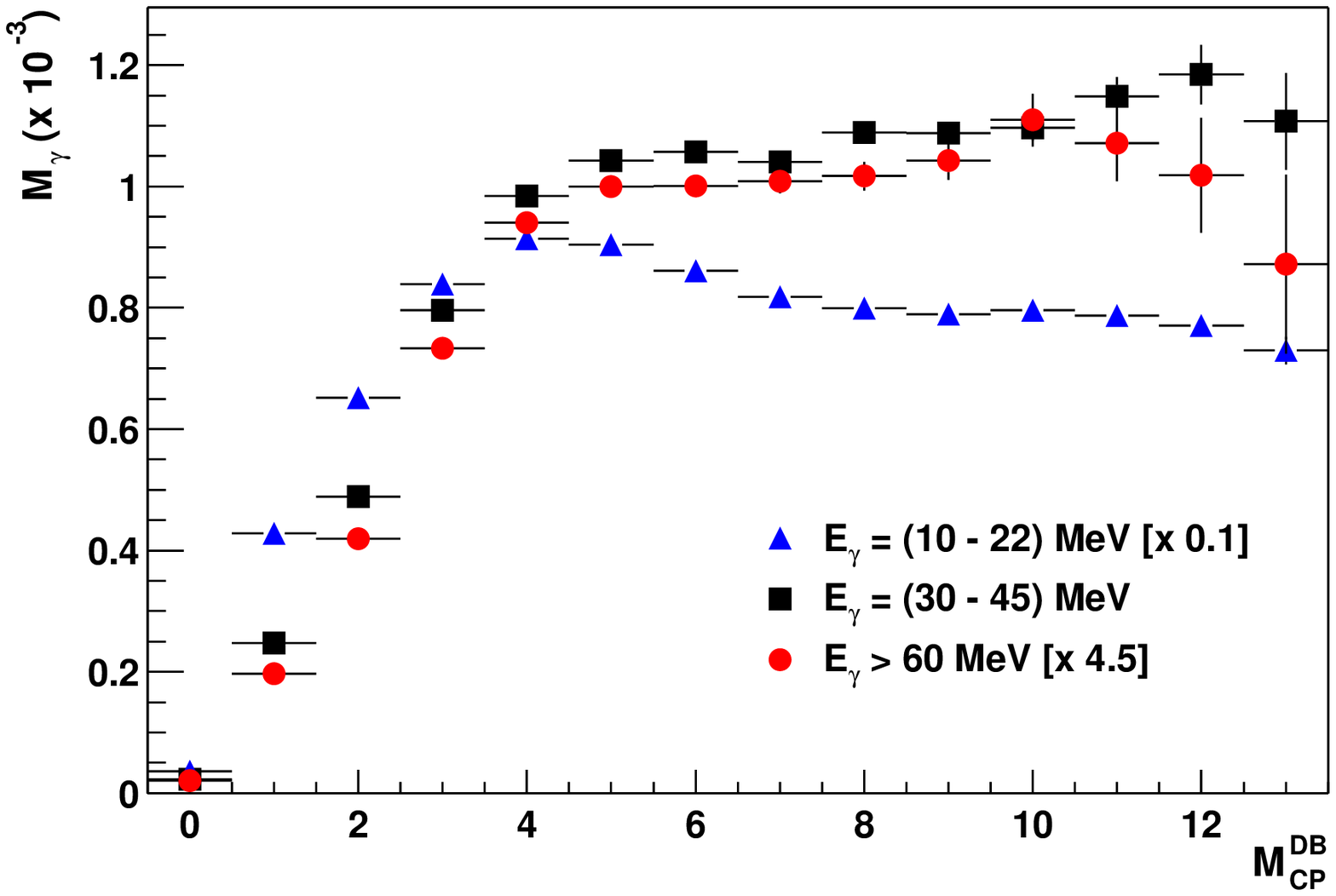} \epsfxsize=7.0cm \epsfbox{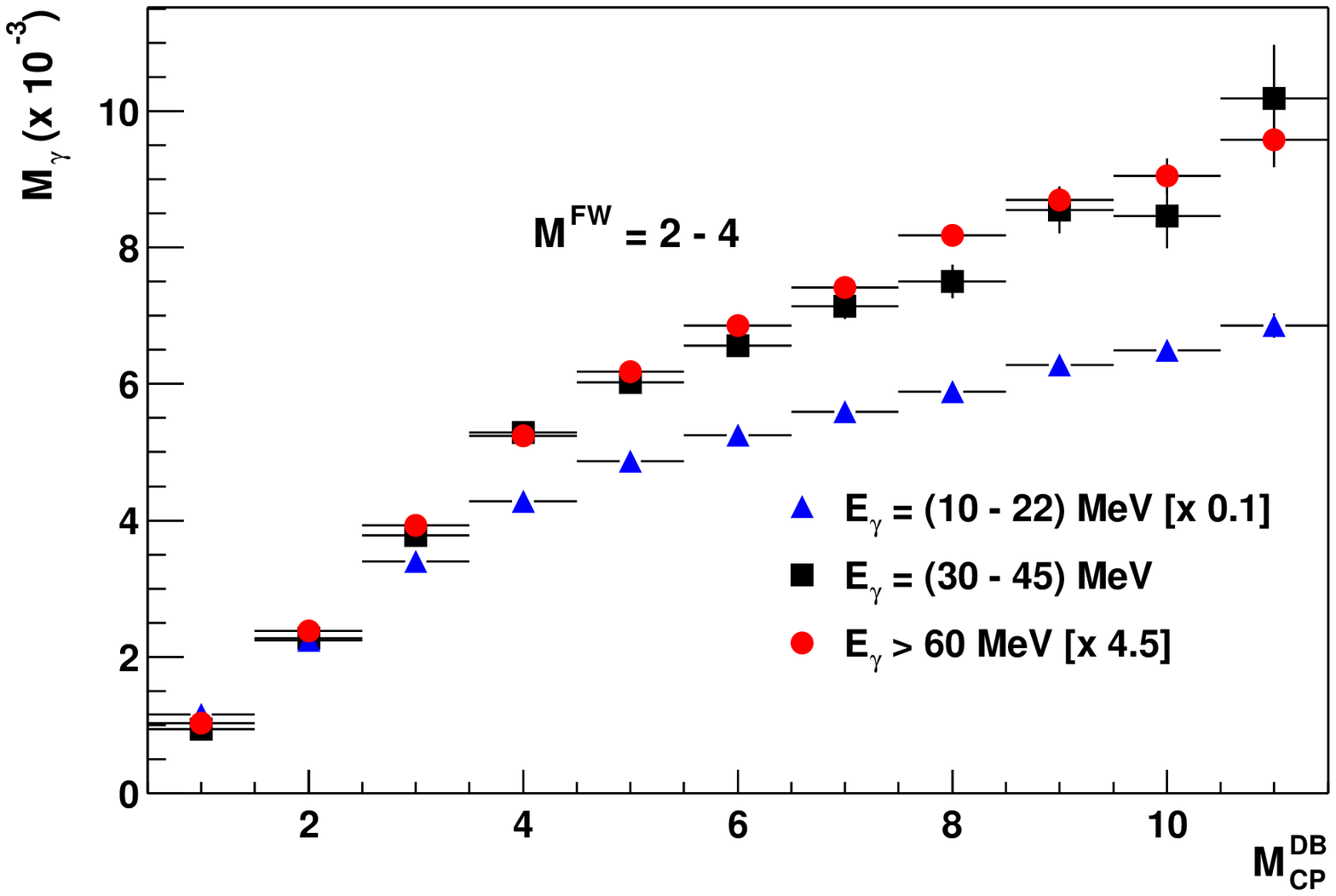}}
 \end{center}
 \caption{a) Photon multiplicity, M$_{\gamma}$, as a function of the charged particle multiplicity measured
 in the DB, M$^{DB}_{CP}$. b) The
 same measurement with the condition of a charged particle multiplicity measured in the FW, M$^{FW}_{CP}$, 
 between 2 and 4, the region of M$^{FW}_{CP}$ in which the photon multiplicity exhibits a maximum.} 
 \label{fig: time.eps}
 \end{figure} 
 \section{Summary and Outlook}

 In summary, we have  analysed the inclusive hard photon spectrum of the 
 $^{129}${Xe}+$^{\rm nat}${Sn} at 50{\it A} MeV reaction. The double-source analysis (a double
 exponential fit with two different inverse slope parameters) shows that the hard-photon distribution is 
 consistent with a direct plus a thermal proton-neutron bremsstrahlung
 contributions, accounting for 78\,\% and 22\,\% of the total hard-photon yield, respectively.
 The measured 
 thermal slope, $E_0^t$, follows  the recently observed linear dependence on the available energy in the 
 nucleus-nucleus center-of-mass $E_{Cc}^{AA}$, 
 confirming a thermal nature of the secondary bremsstrahlung emission. In the analysis of the hard-photon
 angular emission we have observed an isotropic emission of hard photons in the 30 MeV$\,<\mbox{ E$_\gamma$}<\,$40 
 range, consistent with the existence of a thermal
 component coming from second-chance NN collisions which blurs the original elementary anisotropy 
  present in the prompt  direct hard photon emission. In the exclusive hard photon measurements, we have studied
  the dependence of the photon multiplicities on the impact parameter.
 We are presently, carrying out   the same type of analysis of the energy and angular hard-photon spectra
 in reactions in which  multifragmentation is the most probable exit-channel, as well as in peripheral
 collisions, so that we can extract a definite conclusion about thermal hard-photon production in
 multifragment reactions. 

\end{document}